\documentclass{aa}
\usepackage{natbib}
\usepackage{graphicx}
\usepackage{longtable}
\usepackage{amsmath}
\usepackage{url}
\usepackage{txfonts}
\begin{document}
\title{Dimethyl ether in its ground state, $\varv=0$, and lowest two
  torsionally excited states, $\varv_{11}=1$ and $\varv_{15}=1$, in
  the high-mass star-forming region G327.3-0.6}

   \subtitle{}

   \author{S.~E. Bisschop \inst{1}$^,$\inst{2}$^,$\inst{3} \and P. Schilke \inst{4} \and F. Wyrowski\inst{2} \and A. Belloche\inst{2} \and C. Brinch\inst{3}$^,$\inst{1} \and C.~P. Endres\inst{4} \and R. G{\"u}sten\inst{2} \and H. Hafok\inst{2} \and S. Heyminck\inst{2} \and J.~K. J{\o}rgensen\inst{3}$^,$\inst{1} \and H.~S.~P. M{\"u}ller\inst{4} \and K.~M. Menten\inst{2} \and R. Rolffs\inst{2}$^,$\inst{4} \and S. Schlemmer\inst{4}}

   \offprints{S.~E. Bisschop, suzanne@snm.ku.dk}

   \institute{Centre for Star and
     Planet Formation, Natural History Museum of Denmark, University of
     Copenhagen, {\O}ster Voldgade 5-7, Copenhagen K., Denmark DK-1350 \and Max-Planck-Institut f{\"u}r Radioastronomie, Auf dem
     H{\"u}gel 69, 53121, Bonn, Germany \and Centre for Star and Planet Formation, Niels Bohr Institute, Juliane Mariesvej 30, Copenhagen {\O}., DK-2100 \and
     I. Physikalisches Institut, Universit{\"a}t zu K{\"o}ln,
     Z{\"u}lpicher Stra{\ss}e 77, 50937 K{\"o}ln, Germany }

   \date{Received; accepted}

 
  \abstract
  {One of the big questions in astrochemistry is whether complex
    organic molecules are formed in the gas phase after evaporation of
    the icy mantles of interstellar dust grains or at intermediate
    temperatures within these icy mantles. Dimethyl ether
    (CH$_3$OCH$_3$) is one of these species that may form through
    either of these mechanisms, but it is yet unclear which is
    dominant.}
  {The goal of this paper is to determine the respective importance of
    solid state vs. gas phase reactions for the formation of dimethyl
    ether. This is done by a detailed analysis of the excitation
    properties of the ground state and the torsionally excited states,
    $\varv_{11}=1$ and $\varv_{15}=1$, toward the high-mass
    star-forming region G327.3-0.6.}
  {With the Atacama Pathfinder EXperiment 12 m submillimeter
    telescope, we performed a spectral line survey toward G327.3-0.6
    around 1.3, 1.0, and 0.9~mm as well as at 0.43 and 0.37~mm. The
    observed CH$_3$OCH$_3$ spectrum is modeled assuming local thermal
    equilibrium.}
  {CH$_3$OCH$_3$ has been detected in the ground state, $\varv=0$, and
    in the torsionally excited states $\varv_{11}=1$ and
    $\varv_{15}=1$, for which lines have been detected here for the
    first time. The emission is modeled with an isothermal source
    structure as well as with a non-uniform spherical structure. In
    the isothermal case two components at 80 and 100~K are needed to
    reproduce the dimethyl ether emission, whereas an abundance jump
    at 85~K or a model with two abundance jumps at 70 and 100~K fit
    the emission equally well. The emission from the torsionally
    excited states, $\varv_{11}=1$ and $\varv_{15}=1$, is very well
    fit by the same model as the ground state.}
  {For non-uniform source models one abundance jump for dimethyl ether
    is sufficient to fit the emission, but two components are needed
    for the isothermal models. This suggests that dimethyl ether is
    present in an extended region of the envelope and traces a
    non-uniform density and temperature structure. Both types of
    models furthermore suggest that most dimethyl ether is present in
    gas that is warmer than 100~K, but a smaller fraction of 5\%
    --28\% is present at temperatures between 70 and 100~K. The
    dimethyl ether present in this cooler gas is likely formed in the
    solid state, while gas phase formation probably is dominant above
    100~K. Finally, the $\varv_{11}=1$ and $\varv_{15}=1$ torsionally
    excited states are easily excited under the density and
    temperature conditions in G327.3-0.6 and will thus very likely be
    detectable in other hot cores as well.}

   \authorrunning{S.~E. Bisschop et al.}\titlerunning{CH$_3$OCH$_3$ in
     the high-mass star-forming region G327.3-0.6}
   \maketitle
%

\section{Introduction}
\label{intro}

A key stage in high-mass star-formation is the hot molecular core
phase, in which high abundances of complex organic molecules are found
in the inner warm regions \citep[e.g.][]{rodgers2003}. The origin of
complex organic species is still debated, in particular, which species
are formed by solid state reactions in the cooler stages, and which
form in the gas phase after evaporation of icy grain surfaces, when
the young forming high-mass star starts heating up its surroundings
\citep{tielens1997,charnley1992,charnley1995}. CH$_3$OCH$_3$ (dimethyl
ether) is one of these ``hot core molecules'' for which the exact
formation mechanism is not yet clear. It has frequently been detected
toward high-mass star-forming regions and has a typical abundance
of 10$^{-8}$--10$^{-7}$ with respect to H$_2$ \citep[see
e.g.,][]{sutton1995,nummelin2000,schilke2001,comito2005}. It has been
observed towards low-mass star-forming regions as well
\citep{cazaux2003,jorgensen2005b}.

A recent discussion of CH$_3$OCH$_3$ formation is given by
\citet{peeters2006}. Two alternative formation mechanisms are
suggested: grain surface reactions and warm gas phase chemistry. In
the first route, large quantities of CH$_3$OCH$_3$ may be produced by
the combination of radicals formed as the photo-dissociation products
of organic molecules on the surfaces of icy grains
\citep{garrod2008}. Since these radicals are mobile only at
intermediate ice temperatures or higher, CH$_3$OCH$_3$ is formed above
30~K. It evaporates before H$_2$O and CH$_3$OH, due to its inability
to form hydrogen bonds. In laboratory experiments by \citet{oberg2009}
an evaporation temperature of 85--90~K is found. However, the total
number of desorbed molecules is a function of both temperature and
time. Therefore the expected evaporation temperature of CH$_3$OCH$_3$
for more realistic warm-up time scales for star-forming regions, e.g.,
10$^{4}$--10$^{6}$~yr for an envelope warming up from 20 to 200~K
\citep{garrod2008}, is expected to be around 70~K \citep[see
e.g.,][where this time scale effect is demonstrated for CO and
N$_2$]{bisschop2006a}. After evaporation it is destroyed and/or
protonated \citep{peeters2006}. This results in a decreasing abundance
over time and a peak in the CH$_3$OCH$_3$ abundance early in the
evolution of the hot core. The second route is gas phase formation of
CH$_3$OCH$_3$ from CH$_3$OH. This reaction has been measured in the
laboratory by \citet{karpas1989} and occurs in two steps. The first
step is the self-alkylation of protonated methanol:
CH$_3$OH$_2^+~+~{\rm CH}_3{\rm OH} \rightarrow {\rm
  CH}_3$OCH$_4^+~+~{\rm H}_2{\rm O}$. The second step is dissociative
recombination, which leads to the neutral dimethyl
ether. \citet{hamberg2010} report that the dissociative recombination
of CD$_3$OCD$_4^+$ leads to the formation of deuterated dimethyl ether
in about 7\% of the cases. If the same branching ratio holds for the
non-deuterated species, this means that the expected gas phase
abundance ratio CH$_3$OCH$_3$/CH$_3$OH is probably not higher than
$\sim$7\%. The abundance of CH$_3$OCH$_3$ formed in the gas phase is
expected to peak at higher temperatures compared to the grain-surface
formation scenario, since CH$_3$OH only evaporates at around 100~K
\citep{viti2004}. However, low abundances of two orders of magnitude
less of CH$_3$OH are also found in the outer parts of low-mass
protostars and even cold dark clouds
\citep{jorgensen2005c,maret2005,dickens2000}. If this is also the case
for high-mass star-forming regions, it may also lead to the formation
of a small amount of dimethyl ether at low temperatures in the gas
phase. In summary, CH$_3$OCH$_3$ is mainly expected to be present in
the hot core and the emission should be very well modeled with one or
two temperature components. One of the main questions we would like to
answer in this paper is whether this is indeed is the case.

The detailed study of CH$_3$OCH$_3$ is part of a larger project in
which we obtained a large, unbiased, line-survey of the 230, 290, 345,
690 and 810 GHz atmospheric windows with the Atacama Pathfinder
EXperiment\footnote{This publication is based on data acquired with
  the Atacama Pathfinder Experiment (APEX). APEX is a collaboration
  between the Max-Planck-Institut f{\"u}r Radioastronomie, the
  European Southern Observatory, and the Onsala Space Observatory.}
(APEX) telescope for the chemically extremely rich high-mass
star-forming region G327.3-0.6 (Bisschop et al. in prep). The
G327.3-0.6 hot core is located in an active region of high-mass
star-formation, close to two HII regions, and its formation may have
been triggered by an expanding infrared bubble
\citep{minier2009}. North of the ``hot core'' there is a colder cloud
core detected in CO and N$_2$H$^+$ \citep{wyrowski2006}. Recent
observations of mid-J CO-lines by \citet{leurini2012} show that the
molecular emission is extended and suggest that smaller high density
clumps are also present in the region. The aim of our large unbiased
line-survey of the high-mass star-forming region G327.3-0.6 is to make
an inventory of the chemical composition of the ``hot core'' and to
compare their abundances and excitation properties to astrochemical
models of gas and grain-surface chemistry \citep[as for example shown
for ethyl formate and n-propyl cyanide by][]{belloche2009}. Hereby we
aim to get a better understanding of the chemical mechanisms, e.g.,
which species may be formed on grain surfaces and which in the gas
phase. G327.3-0.6 is very suitable for such a study since the emission
lines are strong and narrow. There is relatively little line blending,
which makes it possible to detect many weaker transitions. Next to the
detection of dimethyl ether, a large number of other complex organic
molecules have been detected in this source, such as C$_2$H$_5$CN and
CH$_3$C(O)CH$_3$ \citep[][Bisschop et al. in
prep.]{gibb2000a}. Currently, we have detected 44 molecular species in
this source, 51 isotopologues and 23 vibrationally excited states of
which two are detected for minor isotopologues. Papers in which the
full CHAMP$^+$ and SHeFI surveys will be presented are in preparation.

This paper is structured as follows: Sect.~\ref{obs} discusses the
observations, Sect.~\ref{ana} presents the analysis methods,
Sect.~\ref{results} describes the results of the observations, and
discusses the results of radiative transfer models for the dimethyl
ether emission with either an isothermal or non-uniform density and
temperature structure, in Sect.~\ref{disc} the isothermal and
non-uniform source structure models are compared and the resulting
constraints on the formation mechanisms of CH$_3$OCH$_3$ are discussed,
and finally Sect.~\ref{conclusion} summarizes the main conclusions.

\section{Observations}
\label{obs}

\begin{table}
  \caption{Overview of the observed frequencies, half-power beam widths ($HPBW$) and the telescope beam efficiencies ($B_{\rm eff}$).}\label{freqs}
\begin{center}
\begin{tabular}{lllll}
  \hline
  \hline
$\lambda$& $\nu$ & Freq. range & $HPBW$ & $B_{\rm eff}$\\
mm & GHz & GHz & \arcsec & \\
  \hline
1.3 & 230 & 213--267.5 & 27.1 & 0.81\\
1.0 & 290 & 270--315 & 21.5 & 0.73\\
0.9 & 345 & 335--362 & 18.0 & 0.65\\
0.43 & 690 & 623--714.8 & \phantom{2}8.7 & 0.43$^a$\\
0.37 & 810 & 784--831 \& 843.6-852.8 & \phantom{2}7.6 & 0.35$^b$\\
\hline
\end{tabular}
\end{center}
$^a$ Measured at 661 GHz.
$^b$ Measured at 809 GHz.
\end{table}

G327.3-0.6 was observed in 2008 in the 230, 290 and 345~GHz
atmospheric windows with the Atacama Pathfinder EXperiment (APEX)
located in the Atacama desert in Chile
\citep{gusten2006a,guesten2006}. Additional observations were
performed in August 2009 in the 690 and 810~GHz atmospheric
windows. The coordinates for the G327.3-0.6 hot core are
$\alpha_{\mathrm{J2000}}$=15$^\mathrm{h}$53$^\mathrm{m}$08$\fs$2,
$\delta_{\mathrm{J2000}}$=$-54^\circ$37$\arcmin$06.6$\arcsec$ and the
source has a $V_{\mathrm{lsr}}=-$45~km~s$^{-1}$. We used a distance
of 2.9~kpc for G327.3-0.6 \citep{simpson1990} and a luminosity of
1$\times$10$^5$~L$_\odot$ \citep{wyrowski2006}. In Table~\ref{freqs}
the precise frequency ranges that are covered are given. The
front-ends used were the SHeFI heterodyne receivers, APEX-1 for the
230~GHz window and APEX-2 for the 290 and 345~GHz windows
\citep{vassilev2008}. The 2 times 7-pixel dual channel heterodyne
receiver array, CHAMP$^+$ \citep{kasemann2006}, was employed for the
690 and 810~GHz observations. The Fast Fourier Transform Spectrometer
(FFTS) was the backend for all the SHeFI observations
\citep{klein2006}. It has 8192 spectral channels divided over two
units, each of which has a bandwidth of 1~GHz with a channel spacing
of 122~kHz. The two units are spaced such that there is 100~MHz
overlap, which results in a bandwidth of 1.9~GHz covered per
setting. For the CHAMP$^+$ array the Array Fast Fourier Transform
Spectrometer (AFFTS) was used. It has 2048 channels in each of its two
units, both of which have a bandwidth of 1.5~GHz, a resolution of
732~kHz and an overlap between both units of 180~MHz. All observations
were done in single sideband mode. The telescope pointing was
performed on the continuum of the source itself. The pointing was
checked every two hours for the SHeFI instrument and every hour for
the CHAMP$^+$ instrument and found to be accurate within 2''. The
focus was optimized every few hours or around sun-set and sun-rise and
when large temperature changes occurred. The wobbling mode with a
throw of $\pm$100\arcsec\ was used for the observations.

The main-beam temperatures have been calculated by:

\begin{equation}
T_{\rm MB} = T_{\rm A}^{\ast} \times \frac{F_{\rm eff}}{B_{\rm eff}}
\end{equation}

\noindent where $F_{\rm eff}$ and $B_{\rm eff}$ are the forward and
the main beam efficiencies, respectively. The main beam efficiencies
were based on the measurements at a number of given frequencies: the
SHeFI efficiencies are tabulated in \citet[][]{vassilev2008}, whereas
the efficiencies of the CHAMP$^+$ instrument for the August 2009
observing campaign are given online\footnote{
  \url{http://www3.mpifr-bonn.mpg.de/div/submmtech/heterodyne/champplus/champ_efficiencies.22-08-10.html}}
All values used in this paper are shown in Table~\ref{freqs}. These
efficiencies were scaled to the efficiencies at the observing
frequencies using the Ruze-formula. The half power beam widths
($HPBW$) of APEX \citep{guesten2006,guesten2008} are given by:

\begin{equation}
HPBW = 7.8 \arcsec \frac{800}{\nu {\rm(GHz)}}.
\end{equation}

The rms noise level reached on the $T_{\rm MB}$ scale at 0.244~MHz
resolution was 30--40~mK for the 230 and 290~GHz bands and 45--55~mK
in the 345~GHz window. For the 690~GHz and 810~GHz windows the rms was
100--200~mK and 150--500~mK, respectively, at a resolution of
0.732~MHz.

Before modeling the molecular emission of the source a zeroth order
baseline was subtracted. The baseline was determined based on
emission-free parts of the spectra. In general, the system
temperatures calculated were averaged over the observed range of
1.8~GHz per setting. However, for the CHAMP$^+$ spectra, there are a
few frequency settings where strong atmospheric emission lines are
present. These settings were recalibrated channel-by-channel
off-line. In most cases this reduced the atmospheric features in the
spectra and increased the overall calibration accuracy in the
band. However, the calibration close to these strong atmospheric lines
remains very uncertain and does thus not give reliable quantitative
constraints. We have compared observations of the exact same frequency
taken at different days and from this we estimate that the calibration
uncertainty leads to an overall uncertainty on the line-strengths of
15--20\% for the SheFI data and $\sim$25--30\% for the CHAMP$^+$
data. At 1.3, 1.0, and 0.9~mm the confusion limit is reached for a
significant fraction of the observed frequency ranges, however
sufficient emission free ranges were present to determine a baseline.

\section{Data Analysis}
\label{ana}

This research made use of the {\it myXCLASS} program\footnote{
  \url{http://www.astro.uni-koeln.de/projects/schilke/XCLASS}}
\citep{comito2005}, which accesses the CDMS\footnote{
  \url{http://www.cdms.de}} \citep{muller2001,muller2005} and
JPL\footnote{ \url{http://spec.jpl.nasa.gov}} \citep{pickett1998}
molecular data bases. Additionally, we modeled selected emission lines
of the ground state with the new radiative transfer tool {\it LIME}
\citep{brinch2010}. The line assignments for CH$_3$OCH$_3$ are based
on new measurements of the ground state, $\varv=0$, by
\citet{endres2009} as well as the lowest two torsionally excited
states, $\varv_{11}=1$ and $\varv_{15}=1$ (Endres et al in prep.). The
$\varv_{11}=1$ and $\varv_{15}=1$ states lie 200 and 240~cm$^{-1}$, or
in temperature units 288 and 346~K above the ground state,
respectively. Local thermal equilibrium (LTE) models were constructed
for the emission, in which the source size $\theta$ is given in
$\arcsec$, column density $N_{\rm T}$ in cm$^{-2}$, rotational
temperature $T_{\rm rot}$ in K, and line width $\Delta V$ in
km~s$^{-1}$ were varied to obtain the best-fit model by-eye. We have
estimated the uncertainties by systematically varying all the
parameters in the {\it myXCLASS} models. The model results are shown
together with the uncertainties in Table~\ref{sumres}. The increments
with which the different parameters were varied is equal to half the
uncertainty given in Table~\ref{sumres}.

Although the {\it myXCLASS} models are used to derive the molecular
parameters, rotational diagrams \citep[see][for a discussion of the
method]{goldsmith1999} were used to assess the reliability of the
fits. Rotational diagrams were constructed for the lines of the ground
and torsionally excited states that were not too severely blended,
meaning that these transitions are found to be the major contribution
to the emission feature. Blending with known features of other
molecules has been corrected for by the subtraction of the model for
all molecules minus dimethyl ether from the integrated
line-intensities. Under LTE conditions and assuming the emission is
optically thin the integrated line intensities, $\int T_{\rm MB}dV$ in
K~km~s$^{-1}$ are related to the column density in the upper energy
level, $N_{\rm u}$ in cm$^{-1}$, divided by the degeneracy in the
upper energy level, $g_{\rm u}$, by:
\begin{equation}
\frac{N_{\rm u, thin}}{g_{\rm u}} = \frac{8\pi k \nu^2 \int T_{\rm
MB}dV}{hc^3A_{\rm ul}B} ,\label{roteq1}
\end{equation}
\noindent where $\nu$ is the transition frequency in Hz, $A_{\rm ul}$ is the
Einstein $A$-coefficient in s$^{-1}$. The Einstein $A$-coefficient can be
calculated through:
\begin{equation}
A_{\rm ul} = 1.16395\cdot10^{-20} \nu^3\mu^2S,
\end{equation}
where $\mu$ is the dipole moment in Debye and $S$ is the line
strength. $B$ is the beam-filling factor and is calculated through:
\begin{equation}
B = \frac{\theta_{\rm G327.3-0.6}^2}{\theta_{\rm G327.3-0.6}^2+\theta_{HPBW}^2},
\end{equation}
where $\theta_{\rm G327.3-0.6}$ is the size (FWHM) of the source in
arcsec, which is determined from the isothermal {\it myXCLASS} model
for the optically thick transitions. If the emission for a specific
transition is optically thick the integrated line intensity
additionally has to be multiplied by the correction factor $C_{\tau}$,
which is given by:

\begin{equation}
C_{\tau} = \frac{\tau}{1-e^{-\tau}}, \label{tau}
\end{equation}

\noindent where $\tau$ is the optical depth. Thus:

\begin{equation}
\frac{N_{\rm u, thick}}{g_{\rm u}} = \frac{N_{\rm u, thin}}{g_{\rm u}} C_{\tau} 
\end{equation}

\noindent In this paper the optical depths for the emission lines are
calculated with the isothermal {\it myXCLASS} model. The total {\it
  source-averaged} column density $N_{\rm T}$ can then be computed
from:
\begin{equation}
N_{\rm T} =\frac{N_{\rm u}}{g_{\rm u}} \frac{Q(T_{\rm rot})} 
{e^{-E_{\rm u}/T_{\rm rot}}} ,\label{roteq2}
\end{equation}
where the rotational temperature, $T_{\rm rot}$, should be equal to the
kinetic temperature of the gas under LTE conditions, $Q(T_{\rm rot})$
is the rotational partition function, and $E_{\rm u}$ is the upper
level energy in K. Practically, this means that $T_{\rm rot}$ can be
determined from the slope of the rotational diagram (such as for
example shown in Fig.~\ref{rot_ch3och3-v0}) and $N_{\rm T}$ from the
intercept with the y-axis.
 
The fit to the data has been optimized, using an iterative
approach. First we made a model that fits the data well by-eye and
subsequently a rotation diagram was constructed. The results of this
fit were then used as renewed input until the fit was optimal.

\subsection{Critical density}
\label{crit}

One of the main assumptions in our analysis is that the emission is in
LTE. In the following we will use the critical density, $n_{\rm cr}$
in cm$^{-3}$, to validate this assumption. However, it is necessary to
mention that $n_{\rm cr}$ is not per definition equal to the density
where the excitation temperature is equal to the kinetic
temperature. This depends on for example the frequency of a
transition, and the temperature of the gas
\citep{evans1999,evans1989}. Both higher frequencies and higher
temperatures decrease the densities at which a given transition is
effectively thermalized. At submillimeter wavelengths the
thermalization density is approximately the same as the critical
density. We therefore use the critical density here to estimate the
effective thermalization density. Unfortunately it is not possible to
derive the critical density for transitions of dimethyl ether
directly, since its collisional rates are not known. However, we can
attempt to estimate them, when we assume that the rates for CH$_3$OH
are comparable within an order of magnitude. The critical density for
a transition is calculated by:
\begin{equation}
  n_{\rm cr} = \frac{A_{\rm ul}}{\gamma\ }. \label{criteq}
\end{equation}

\noindent Here $\gamma$ is the collisional rate in
cm$^3$~s$^{-1}$. The collisional rates for CH$_3$OH are
$\sim$10$^{-11}$--10$^{-10}$~cm$^3$~s$^{-1}$ \citep{pottage2004}. The
values for the Einstein {\it A}-coefficients of CH$_3$OCH$_3$ range
from 10$^{-6}$~s$^{-1}$ up to 4.0$\times$10$^{-3}$~s$^{-1}$ at the
highest frequencies and up to 3.0$\times$10$^{-4}$~s$^{-1}$ for
frequencies below 370~GHz. This means that $n_{\rm cr}$ can be as high
as 10$^7$--10$^8$~cm$^{-3}$ for the lines with the highest
line-strengths assuming the collisional rate is
10$^{-11}$~cm$^3$s$^{-1}$. However, as previously mentioned the
effective density at which a transition is thermalized is likely to be
different from the critical density. Multiple effects can play a role,
e.g., for transitions from moderately to highly excited states the sum
over all collisional rates from the upper state need to be considered
and not only the one corresponding to the radiative transition. High
optical depths can also lower the density at which a transition is
thermalized. Additionally, a strong infrared radiation field can
thermalize the populations to the radiation field irrespective of the
density. However we do not expect the latter to be a problem for
dimethyl ether in this source (see Sect.~\ref{excited}). When effects
like high optical depth or the inclusion of all collisional rates are
taken into account, the density at which the molecular emission
reflects the kinetic temperature is one or two orders of magnitude
lower than the critical density, i.e.,
10$^{5}$--10$^{6}$~cm$^{-3}$. Since we used a rather conservative
estimate for the collisional rate, the effective thermalization
density may be lower still. This means that most transitions should
easily be thermalized at densities of 10$^{6}$--10$^{7}$~cm$^{-3}$ we
expect at temperatures from 70~K and higher from the model by
\citet{rolffs2011}.

\subsection{Dust optical depth}
\label{dust}

The observations described in this paper cover a large frequency range
and it is therefore important to obtain a model that is consistent
over all frequencies. The LTE models constructed with the {\it
  myXCLASS} software systematically overestimate the line-strength for
the CHAMP$^+$ part of the survey. This is likely due to the dust
optical depth at higher frequencies, which absorbs part of the
radiation emitted by the molecules \citep{mezger1990}. This can be
corrected for by estimating the total column density of hydrogen,
$N_{\rm H}$, from which the dust optical depth is calculated such that
the relative intensities over all frequencies match. $N_{\rm H}$ is
given by $N({\rm H})+ 2N({\rm H_2})$ in cm$^{-2}$. The dust optical
depth, $\tau_{\rm d}$, is calculated by:
\begin{equation}
\tau_{\rm d} = N_{\rm H}\ \sigma_\lambda^{H},
\end{equation}

\noindent The dust collisional cross section ($\sigma_\lambda^{\rm
  H}$) is given by:

\begin{equation}
\sigma_\lambda^{\rm H} = 7\times10^{-21}\ \frac{Z}{Z_\odot}\ b \lambda_{\rm \mu m}^{-2} ,
\end{equation}

\noindent for $\lambda > 100~\mu$m. The metallicity for G327.3-0.6,
$\frac{Z}{Z_\odot}$, is assumed to be equal to that in the galactic
center, i.e. 2, and $b$ is an adjustable factor, that takes grain
properties into account. The {\it myXCLASS} program assumes a value of
3.4 for deeply embedded IR sources where the grains are expected to be
coated with thin ice layers \citep{rengarajan1984}. In comparison, a
factor of 1.9 is used for grains without ice layers. The latter may be
more correct for the CH$_3$OCH$_3$ emission that is arising from
warmer gas. Furthermore in the {\it myXCLASS} program, dust is
considered to be present in a foreground layer and corrected for with
the factor e$^{-\tau_{\rm d}}$. For consistency we use the same
correction factor for the rotational diagram treatment. However, when
the dust is present in the source itself molecular line emission is
expected to be attenuated with the factor $(1-e^{-\tau_{\rm
    d}})/\tau_{\rm d}$. The combination of the effect of the dust
grain properties and foreground/in situ treatment could lead to an
underestimation of the value for $N_{\rm H}$ with a factor 4--5 in
addition to the uncertainty of about $\sim$50\% on the value derived
from the CHAMP$^+$ observations with {\it myXCLASS}. However, with the
additional uncertainties on the source structure and the dust
composition it is accurate enough for our purposes. The
line-intensities over the full frequency range covered in this paper
are corrected for by the factor e$^{-\tau_{\rm d}}$. The optimal value
for $N_{\rm H}$ is derived by assuming the model for dimethyl ether
found at the lower frequencies also to hold at the higher frequencies
(See Sect.~\ref{groundsec}). Typical values for $\tau_{\rm d}$ are
0.08 (230~GHz), 0.13 (290~GHz), 0.19 (345~GHz), 0.76 (690~GHz) and
1.04 (810~GHz).

\section{Results}\label{results}

\subsection{The CH$_3$OCH$_3$ ground state,  $\varv=0$}
\label{groundsec}

\begin{figure*}
\includegraphics[width=16cm]{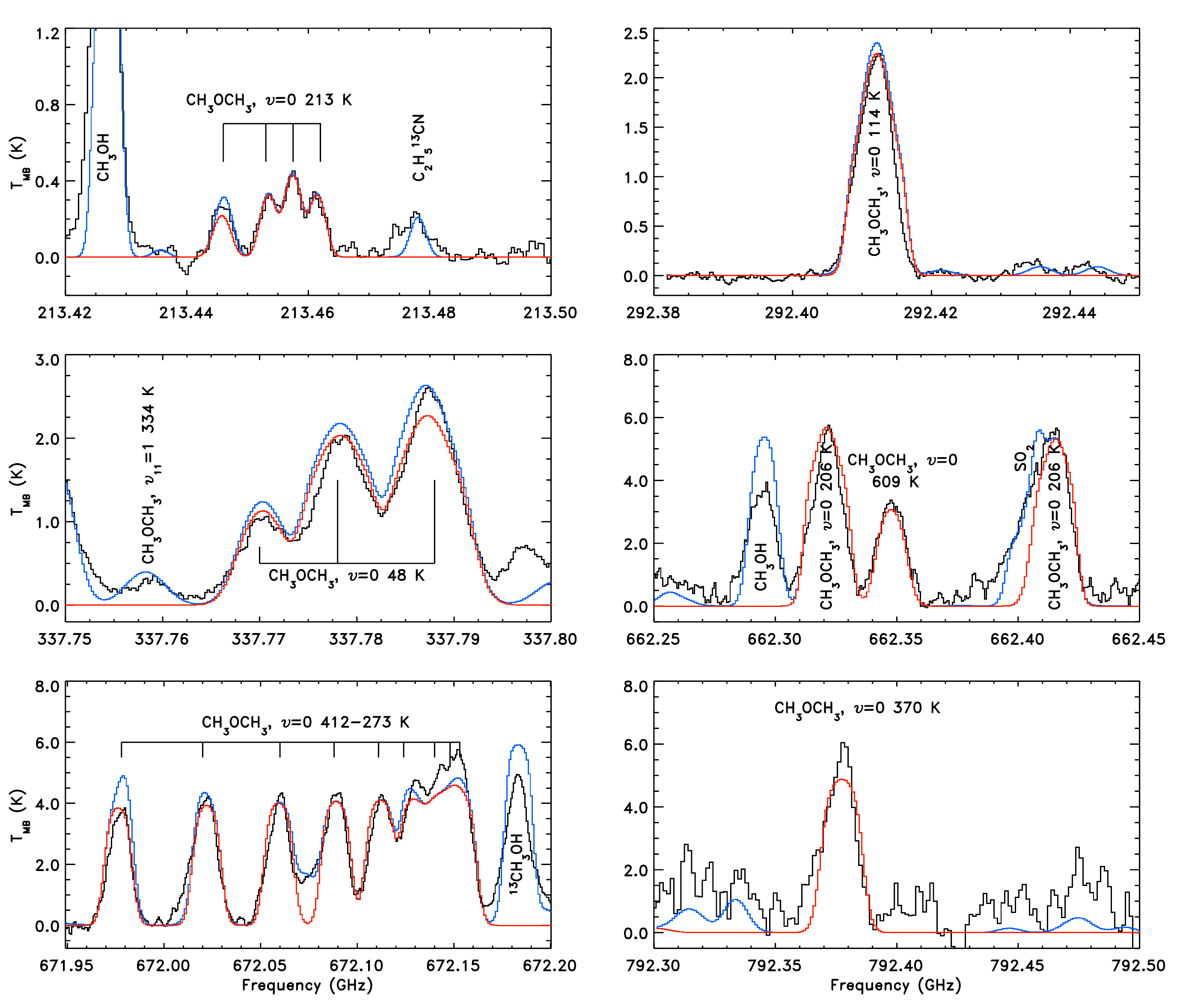}
\caption{Selected emission lines for ground-state CH$_3$OCH$_3$,
  $\varv=0$, in all atmospheric windows. The observed spectrum is
  displayed in black, the {\it myXCLASS} isothermal model for
  CH$_3$OCH$_3$, $\varv=0$, in red and the {\it myXCLASS} isothermal
  model for all assigned species in blue. The major transitions of
  CH$_3$OCH$_3$, $\varv=0$, and other detected species are indicated
  in the plot as well as the values for $E_{\rm u}$ for the
  transitions of CH$_3$OCH$_3$, $\varv=0$.}\label{gsspec}
\end{figure*}

\begin{figure*}
  \resizebox{18cm}{!}{\includegraphics{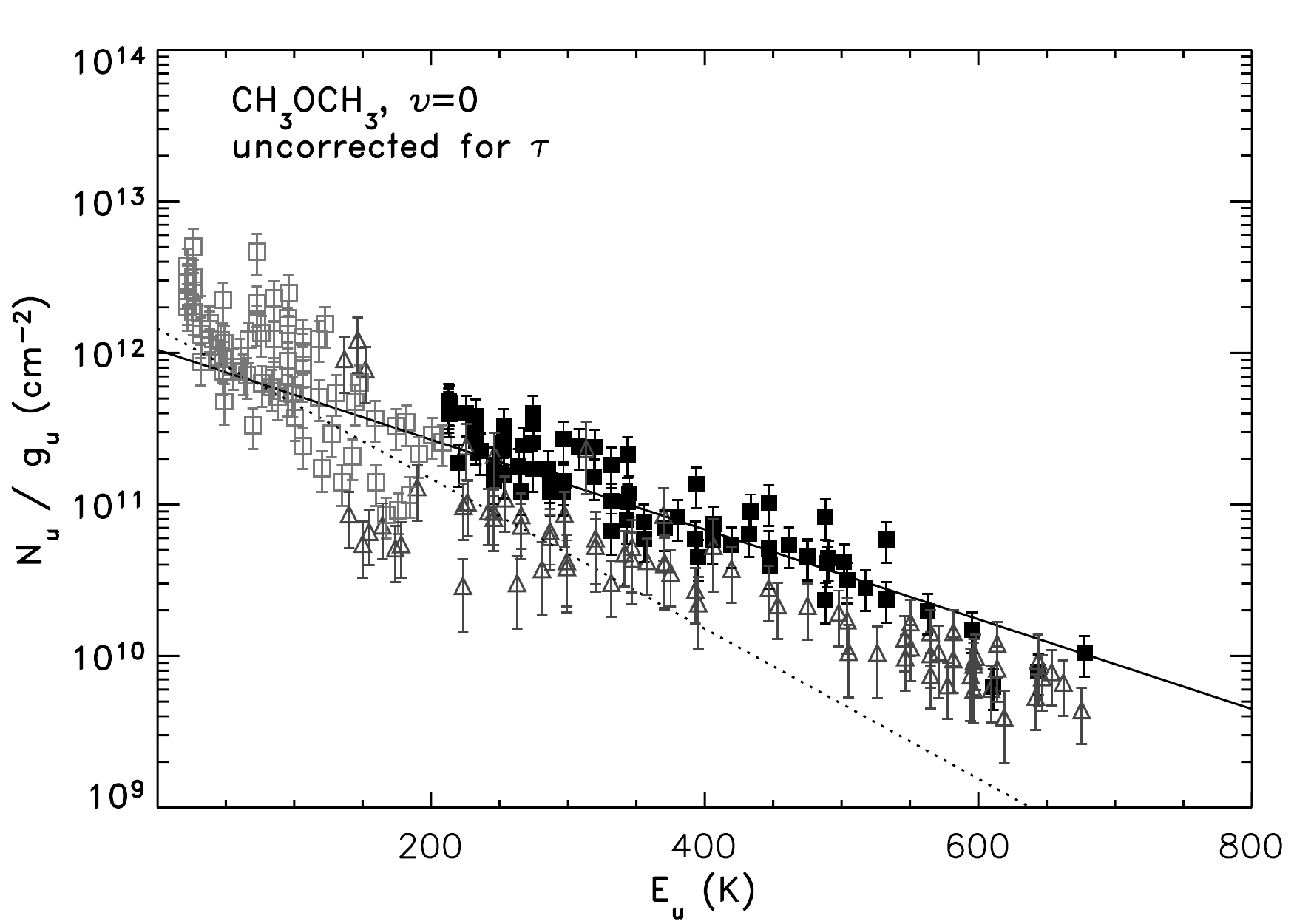}\includegraphics{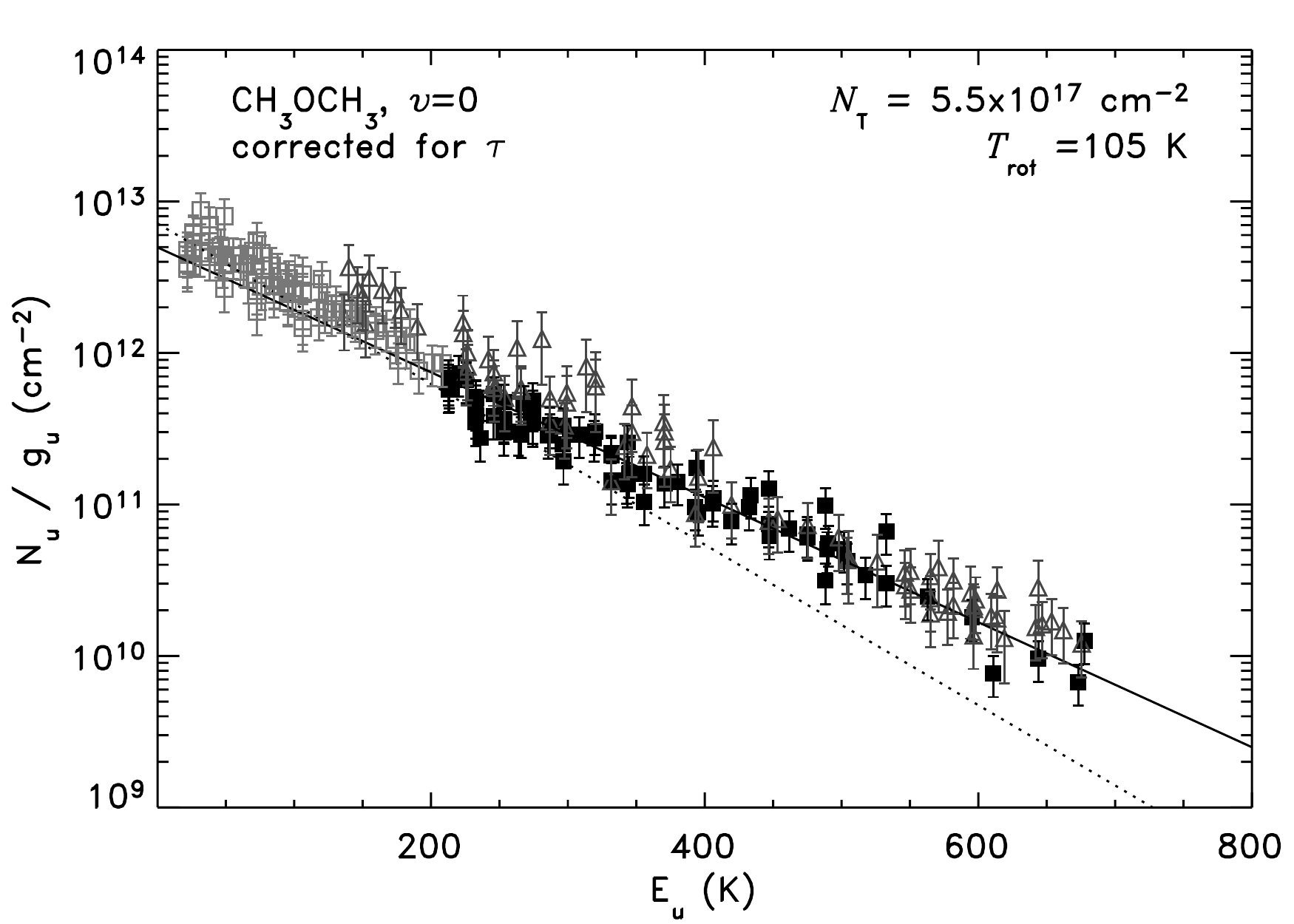}}
  \caption{Rotational diagram for CH$_3$OCH$_3$, $\varv=0$,
    uncorrected for the predicted optical depth (left) and corrected
    for the optical depth (right), in all other aspects the data is
    treated in the same way, as well as that the same symbols and
    line-styles are used in both panels. Line-emission from other
    species, derived from an isothermal model of all identified
    species minus dimethyl ether, has been subtracted from the
    dimethyl ether emission in both figures. The filled squares
    ($\blacksquare$) indicate the lines with $E_{\rm u}>$~200~K and
    open squares ($\Box$) lines with $E_{\rm u}<$~200~K, detected with
    SHeFI in the 230, 290 and 345~GHz windows. The open triangles
    ($\triangle$) indicate CHAMP$^+$ data. The solid line shows the
    least squares-fit to the $E_{\rm u}>$~200~K data and the dotted
    line the least-square fit to the $E_{\rm u}<$~200~K
    transitions. The dust absorption has been corrected for. $N_{\rm
      T}$ for the fit to the transitions with $E_{\rm u}~>$~200~K is
    5.5$\times$10$^{17}$~cm$^{-2}$ and $T_{\rm rot}$ is 105~K for a
    source size of 2.6\arcsec. For the transitions with $E_{\rm
      u}~<$200~K these values are 5.0$\times$10$^{17}$~cm$^{-2}$ and
    82~K, respectively for a source size of
    3.2\arcsec.} \label{rot_ch3och3-v0}
\end{figure*}

Emission lines for the ground state are detected in all atmospheric
windows, more specifically 337 in the SHeFI bands and 136 at the
frequencies observed with CHAMP$^+$. Examples of lines detected for
each band are shown in Fig.~\ref{gsspec}. Naturally a large number of
lines is blended with transitions of other species, which results in a
number of 250 features, for which CH$_3$OCH$_3$ is the main
contributor (an overview of the CH$_3$OCH$_3$, $\varv=0$ transitions is
given in the appendix in Table~\ref{detfreqv0}). As described in
Sect.~\ref{ana}, the model for the emission of all other assigned
features is taken into account in the isothermal {\it myXCLASS}
model. The values for $E_{\rm u}$ for the detected transitions range
from 21.7 to 677.5~K, and are very evenly spread in energy. The
optical depths for detected transitions range from 0.056 to 26.

\begin{table}
  \caption{Resulting best by-eye fits for the isothermal {\it myXCLASS} model. The estimated uncertainties on the different parameters are given in brackets.}\label{sumres}\centering
\begin{tabular}{llll}
  \hline
  \hline
$\theta$& $T_{\rm rot}$ & $N_{\rm T}^a$ & $\Delta V$ \\
\arcsec & K & cm$^{-2}$ & km s$^{-1}$\\
\hline
\multicolumn{4}{l}{One-component model}\\
\hline
3.2(0.2) & 90(5) & 8.0$\times$10$^{17}$(0.5) & 4(0.3)\\
\hline
\multicolumn{4}{l}{Two-component model}\\
\hline
2.6(0.2) & 100(5) & 1.5$\times$10$^{18}$(0.3) & 4(0.3) \\
3.2(0.2) & \phantom{1}80(5) & 7.5$\times$10$^{16}$(2.0) & 4(0.3)  \\
\hline
\multicolumn{4}{l}{$^aN_{\rm T}$ is the source averaged column density calculated using Eq.~\ref{roteq2}.}
\end{tabular}
\end{table}

We have attempted to fit the data with both a one-component and
two-component isothermal model (see Sect.~\ref{ana}). In general the
two components fit the data better by-eye, since it is difficult to
model both the intensities of the optically thick transitions with
excitation temperatures above and below 200~K accurately with only one
component. The optically thick lines with the lowest excitation
temperatures of $\sim$30~K namely suggest a significantly larger
source size compared to the optically thick lines of $\sim$200~K. To
get an idea of how well both models fit we have calculated a reduced
$\chi^2$ for both isothermal models for the same transitions as
modeled with the non-uniform density model discussed in
Sect.~\ref{limeana}. The parameters for both models, as well as the
reduced $\chi^2$ are shown in Table~\ref{sumres}. The two-component
isothermal model has a reduced $\chi^2$ of 1.34 vs. 2.08 for the
one-component model. This result may suggest that there are two
physical components, but could also indicate that the dimethyl ether
emission arises from a region with a non-uniform density and/or
temperature structure. The same is seen in the CHAMP$^+$ data. Here
two different values for $N_{\rm H}$ are needed to correct for the
dust optical depth (see Sect.~\ref{dust} for an explanation). This is
also illustrated in Fig.~\ref{rot_ch3och3-v0}, where the CHAMP$^+$
data (shown as $\triangle$) is corrected for the $\tau_{\rm d}$ found
for the highest excitation lines for a hydrogen column density of
$2\times$10$^{24}$~cm$^{-2}$. This value is consistent within the
uncertainties mentioned in Sect.~\ref{dust} with ${N}_{T \rm >100~K}$
of 4.5$\times$10$^{24}$~cm$^{-2}$ derived for the models by
\citet{rolffs2011} from the 870~$\mu$m dust emission, in particular
since the value found with {\it myXCLASS} (see Sect.~\ref{dust})
likely underestimates the actual value. The CHAMP$^+$ detections are
well fit at the highest excitation energies, but that the excitation
levels below 200~K actually are corrected too much and lie above the
fit and most of the transitions detected with SheFI (shown in
Fig.~\ref{rot_ch3och3-v0} as $\Box$). These lower energy CHAMP$^+$
detections thus seem to require a lower dust optical depth and
corresponding hydrogen column density ($N_{\rm
  H}=$1$\times$10$^{24}$~cm$^{-2}$) to match the lower frequency
data. This could be explained by dimethyl ether gas that is present in
an extended region with a temperature and density gradient or that has
two different abundance components. Both scenarios may well cause
dimethyl ether to ``see'' different dust columns dependent on the
location in the envelope of the star-forming region (see also
Sect.~\ref{disc}).

The two-component isothermal {\it myXCLASS} model fit to the data is
shown in the rotation diagram in Fig.~\ref{rot_ch3och3-v0} for the
ground state. On the left the rotation diagram is shown without the
correction for optical depth and on the right with the correction for
the optical depth derived from the model. Least-square fits of the
transitions are shown for both $E_{\rm u}$ values above 200~K (solid
line) and below 200~K (dotted line). The column density for the fit to
the transitions with $E_{\rm u}~>$~200~K is
5.5$\times$10$^{17}$~cm$^{-2}$ and the rotational temperature is 105~K
for a source size of 2.6\arcsec. For the transitions with $E_{\rm
  u}~<$200~K these values are 5.0$\times$10$^{17}$~cm$^{-2}$ and 82~K,
respectively for a source size of 3.2\arcsec. The difference between
the column densities from the rotational diagram and the two
components given in Table~\ref{sumres} are due to the fact that the
two temperature components of the isothermal model are both
contributing to the emission of all transitions, and the least-square
column densities are therefore higher. The comparison between the two
rotation diagrams clearly shows that a large fraction of the
``scatter'' is due to the optical depth of the emission lines. The
rotational temperature derived in the uncorrected rotational diagram
is furthermore higher and the column density about an order of
magnitude lower compared to that corrected for optical depth. Some
scatter remains for the corrected data which is most likely due to the
presence of line-blends with unidentified features, as well as
uncertainties in the baseline level due to line-confusion. 

As discussed in Sect.~\ref{crit}, it is possible (though not likely)
that some transitions for dimethyl-ether are excited under non-LTE
conditions. If we remove those lines for which this is most probable,
the fit to the data is not affected significantly, and most of these
transitions are very well-fit by the LTE model. Thus we conclude that
LTE indeed is a good approximation for dimethyl ether in this source.

\subsection{The CH$_3$OCH$_3$ torsionally excited states
  $\varv_{11}=1$ and $\varv_{15}=1$}\label{excited}

\begin{figure*}
\includegraphics[width=16cm]{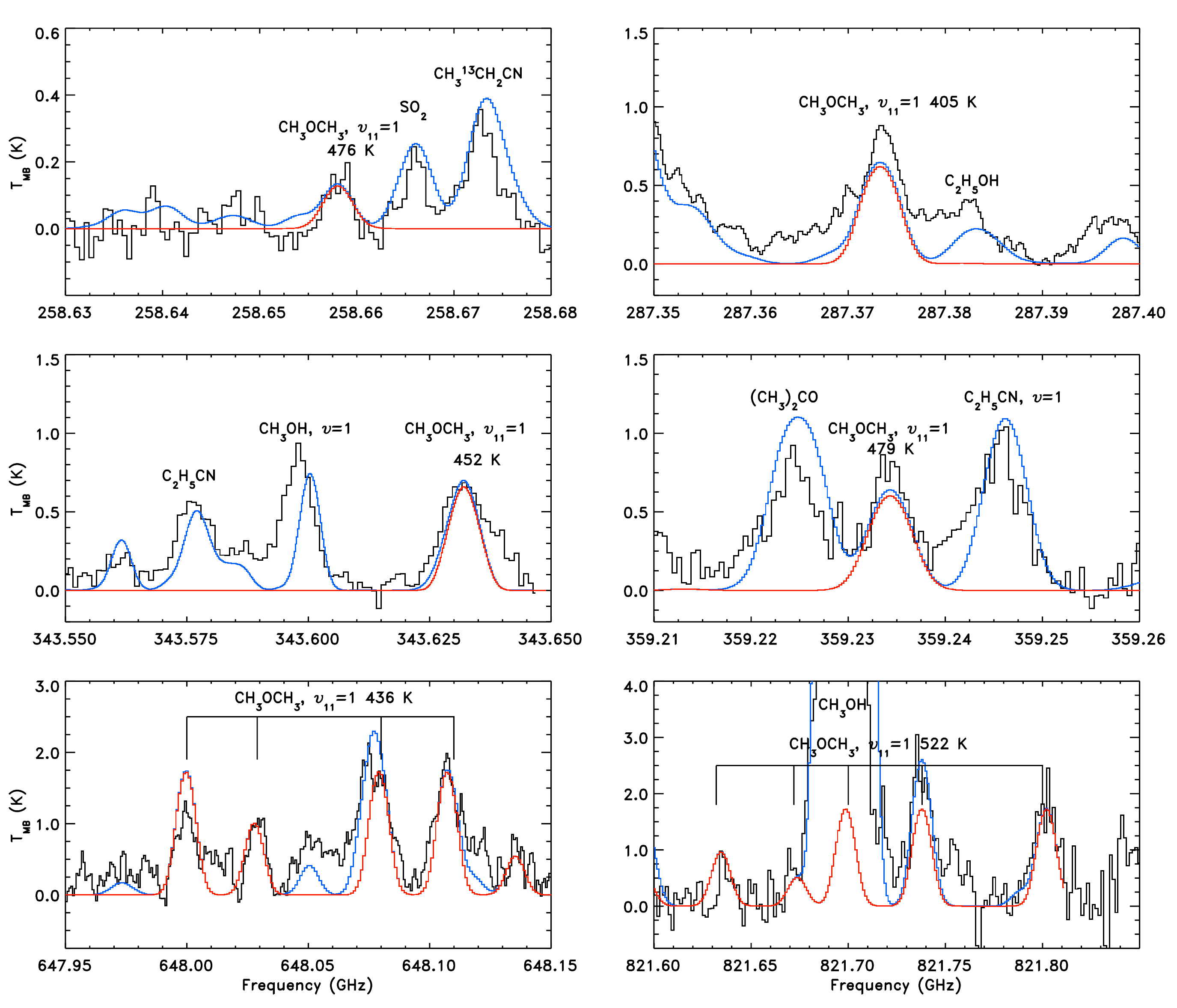}
\caption{Selected emission lines for CH$_3$OCH$_3$, $\varv_{11}=1$, in
  all atmospheric windows. The observed spectrum is displayed in
  black, the {\it myXCLASS} isothermal model for only CH$_3$OCH$_3$,
  $\varv_{11}=1$, in red and the model for all assigned species in
  blue. Transitions of CH$_3$OCH$_3$, $\varv_{11}=1$, and other
  detected species are labeled in the plot as well as the values for
  $E_{\rm u}$ for CH$_3$OCH$_3$, $\varv_{11}=1$.}\label{v11spec}
\end{figure*}

\begin{figure*}
\includegraphics[width=16cm]{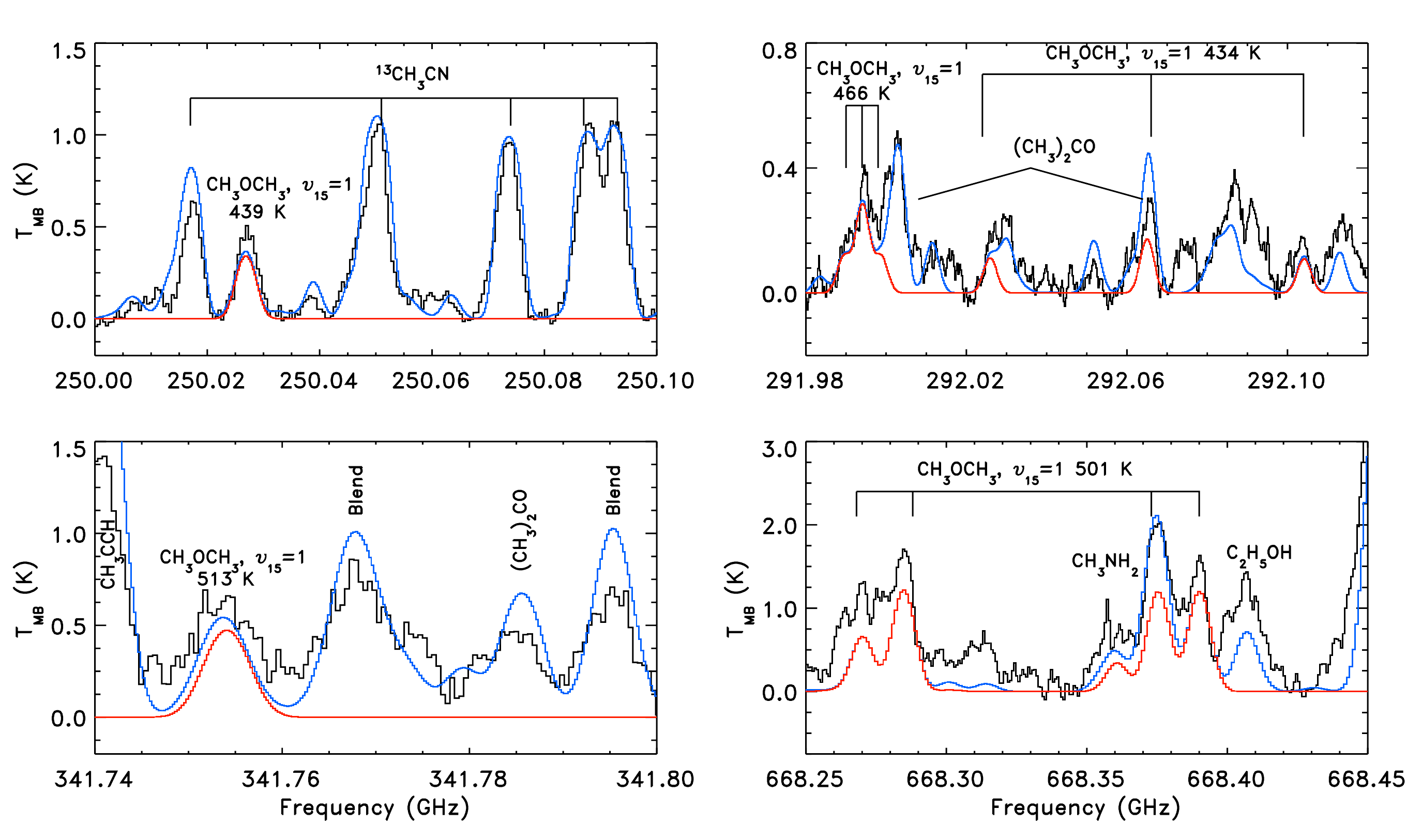}
\caption{Selected emission lines for CH$_3$OCH$_3$, $\varv_{15}=1$ in
  four atmospheric windows. The observed spectrum is displayed in
  black, the {\it myXCLASS} isothermal model for only CH$_3$OCH$_3$,
  $\varv_{15}=1$, in red and the model for all assigned species in
  blue. Transitions of CH$_3$OCH$_3$, $\varv_{15}=1$, and other
  detected species are labeled in the plot as well as the values for
  $E_{\rm u}$ for CH$_3$OCH$_3$, $\varv_{15}=1$.}\label{v15spec}
\end{figure*}

Emission from the $\varv_{11}=1$ torsionally excited state is detected
in all atmospheric windows studied here. The $\varv_{15}=1$ state is
detected in all atmospheric windows except for the highest (810~GHz)
window. The number of lines present in the observations for
$\varv_{11}=1$ with the SHeFI receiver is 117 and 84 with the
CHAMP$^+$ receiver. For $\varv_{15}=1$ these numbers are 59 and 39,
respectively. Unfortunately, some of these features are very strongly
blended with transitions of other species so that there are only 101
and 42 ``clean'' features that can reliably be used for the analysis
of $\varv_{11}=1$ and $\varv_{15}=1$, respectively (see for an
overview of these assignments Tables \ref{detfreqv11} and
\ref{detfreqv15} in the appendix). In Figs.~\ref{v11spec} and
\ref{v15spec} selected transitions for both states are shown for all
atmospheric windows in which they are detected. The torsionally
excited states states lie $\sim$288~K ($\varv_{11}=1$) and 346~K
($\varv_{15}=1$) above the ground state. Lines with $E_{\rm u}$ up to
550~K have been detected. The smaller energy range covered compared to
the ground state, means that the column densities and rotational
temperatures for $\varv_{11}=1$ and $\varv_{15}=1$ lines are less
well-constrained. However, their spectra are very well modeled with
the same LTE model as the ground state (see
Sect.~\ref{groundsec}). Interestingly the $\varv_{11}=1$ torsionally
excited state is infrared inactive while $\varv_{15}=1$ is infrared
active. If dimethyl ether would strongly interact with the infrared
radiation field, a difference between the excitation of the ground
state and the torsionally excited states or between the two
torsionally excited states would be expected. Since this clearly is
not the case, we conclude that for this specific molecule it is not
necessary to include or correct for an infrared radiation field in the
models.

\begin{figure*}\centering
\resizebox{\hsize}{!}{\includegraphics[width=9.2cm]{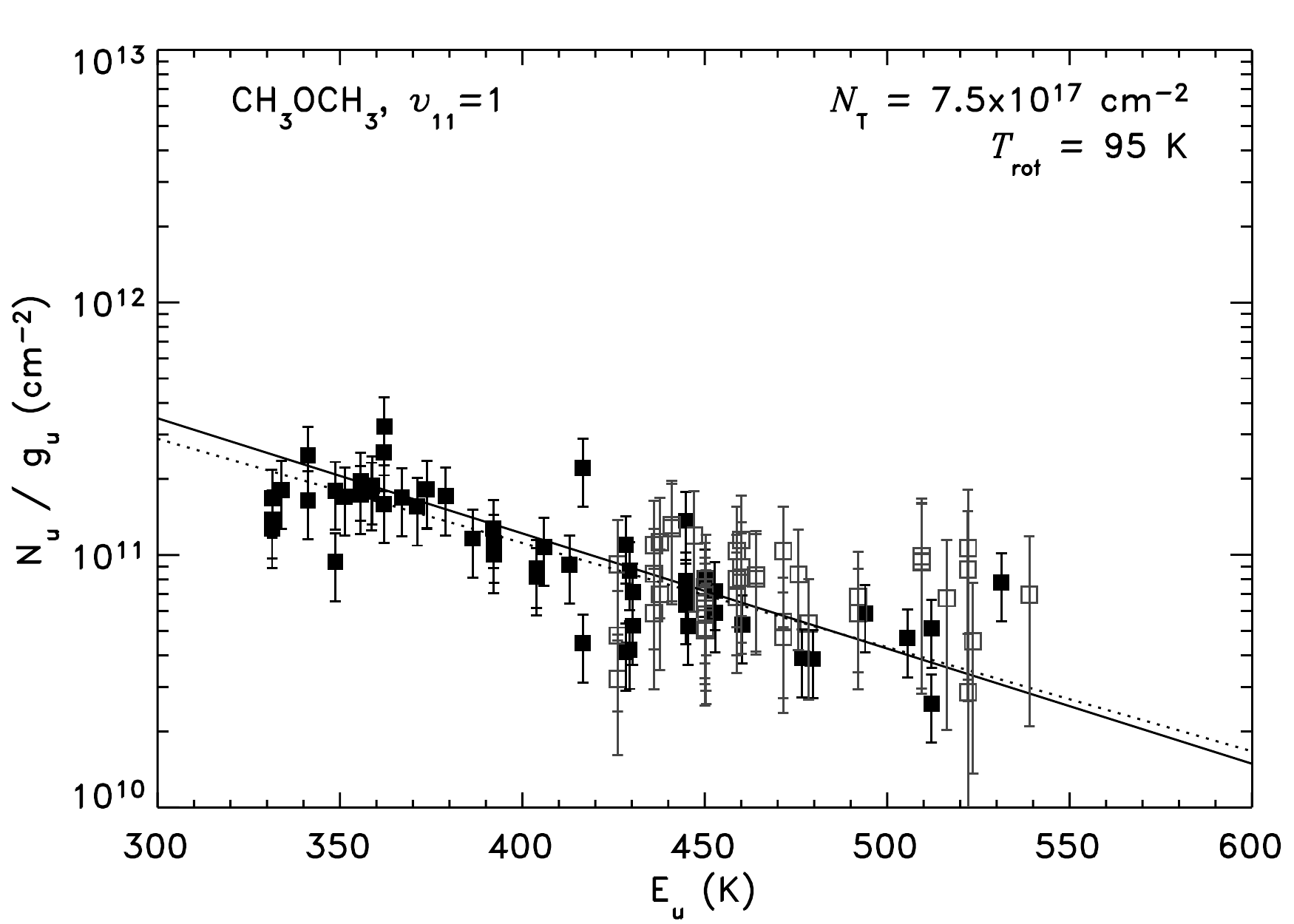}\includegraphics[width=9.2cm]{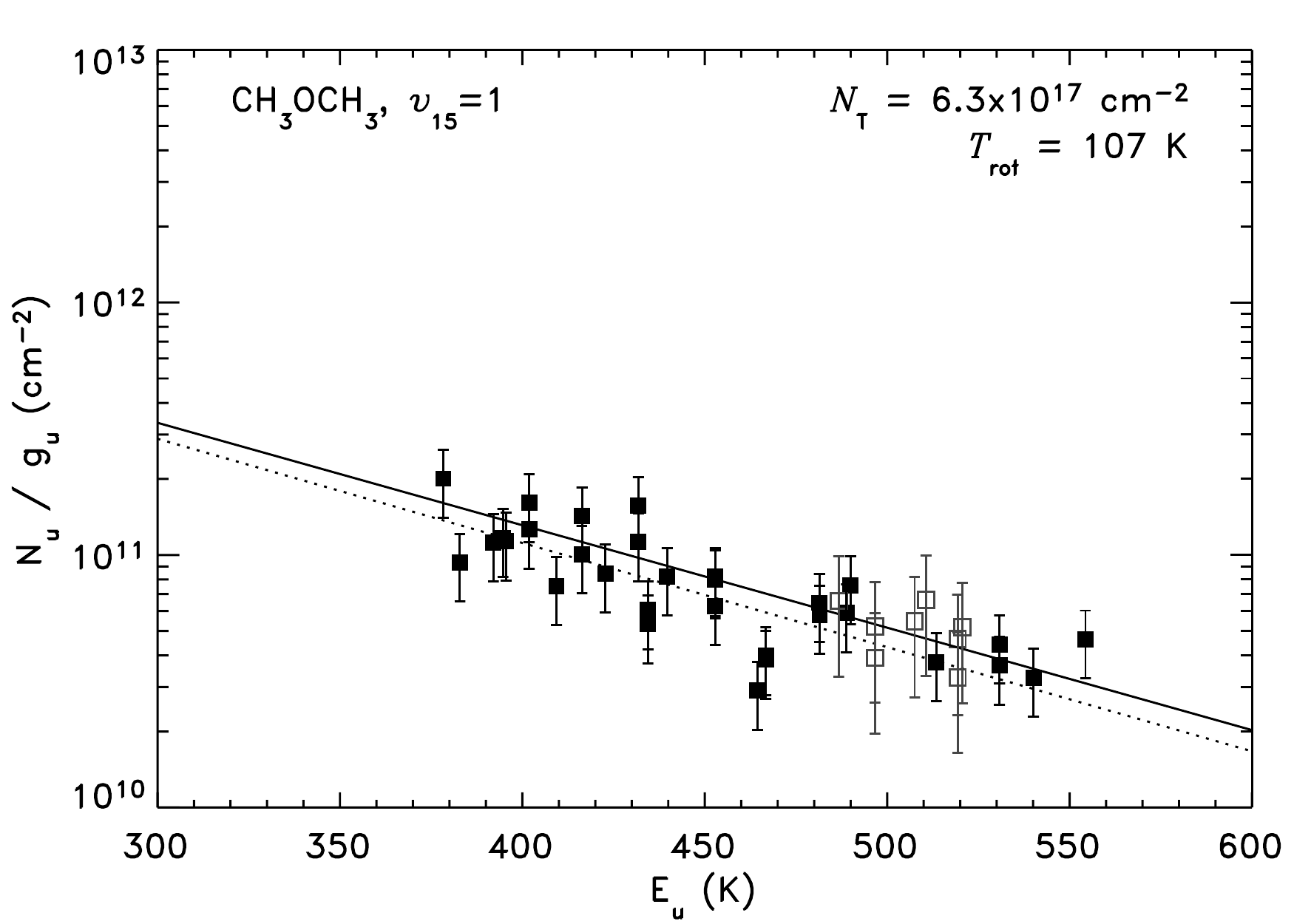}}
\caption{Rotational diagrams for the torsionally excited states,
  $\varv_{11}=1$ (left) and $\varv_{15}=1$ (right), of dimethyl
  ether. The filled squares ($\blacksquare$) indicate transitions
  detected with the SHeFI instrument, the open squares ($\Box$)
  detections with the CHAMP$^+$ instrument. The solid line represents
  the fit to the torsionally excited states and the dotted line
  indicates the fit to the ground state transitions with $E_{\rm
    u}>$~200~K. The least-square fit to the torsionally excited state,
  $\varv_{11}=1$ gives a resulting $N_{\rm T}$ of
  7.5$\times$10$^{17}$~cm$^{-2}$ and a $T_{\rm rot}$ of 95~K. For the
  torsionally excited state, $\varv_{15}=1$, these values are
  6.3$\times$10$^{17}$~cm$^{-2}$ and 107~K,
  respectively.} \label{rot_ch3och3-v11}
\end{figure*}

In Fig.~\ref{rot_ch3och3-v11} the rotational diagrams are shown for
the $\varv_{11}=1$ and $\varv_{15}=1$ states. The least-square fits
are indicated by the solid black line and a fit to the high energy
transitions of the ground state is indicated with the dotted line. The
least-square fit to the torsionally excited state, $\varv_{11}=1$
gives a resulting column density of 7.5$\times$10$^{17}$~cm$^{-2}$ and
a rotational temperature of 95~K. For the torsionally excited state,
$\varv_{15}=1$, these values are 6.3$\times$10$^{17}$~cm$^{-2}$ and
107~K, respectively. The consistency between the model for the ground
state and the torsionally excited states is not just clear from the
isothermal model fits shown in Figs.~\ref{v11spec} and \ref{v15spec},
but is also apparent in these rotational diagrams
(Fig.~\ref{rot_ch3och3-v11}). Due to the lower optical depths for the
transitions of the excited states one could expect that the critical
densities for the lines from the $\varv_{11}=1$ and $\varv_{15}=1$
states and thus the densities where LTE is a good approximation are
higher than for the ground state. Nonetheless they are very well fit
by the same model, confirming our conclusion in Sect.~\ref{crit} that
LTE is a good assumption for dimethyl ether toward the G327.3-0.6
star-forming region.

The optical depths in the isothermal models for the torsionally
excited states range from 0.05--0.655 and 0.038--0.387 for
$\varv_{11}=1$ and $\varv_{15}=1$, respectively. The line intensities
are very close to what is expected based on the ground state and no
lines that are expected to be there are missing. We are therefore
confident, that we have detected the $\varv_{11}=1$ and $\varv_{15}=1$
states here for the first time in space and expect that their
transitions will easily be detected in other hot core sources in the
future.

\subsection{A non-uniform density and temperature model} \label{limeana}

In Sect.~\ref{groundsec} we show that two abundance components
  are needed to model the dimethyl ether emission with an isothermal
  model. This may either be due to two physical components or to the
  fact that the dimethyl ether emission arises from an extended region
  in the envelope. We now want to test these scenarios for a
spherical model for the source structure with a physically more
realistic temperature and density profile.

\begin{figure*}\centering
\includegraphics[width=18cm]{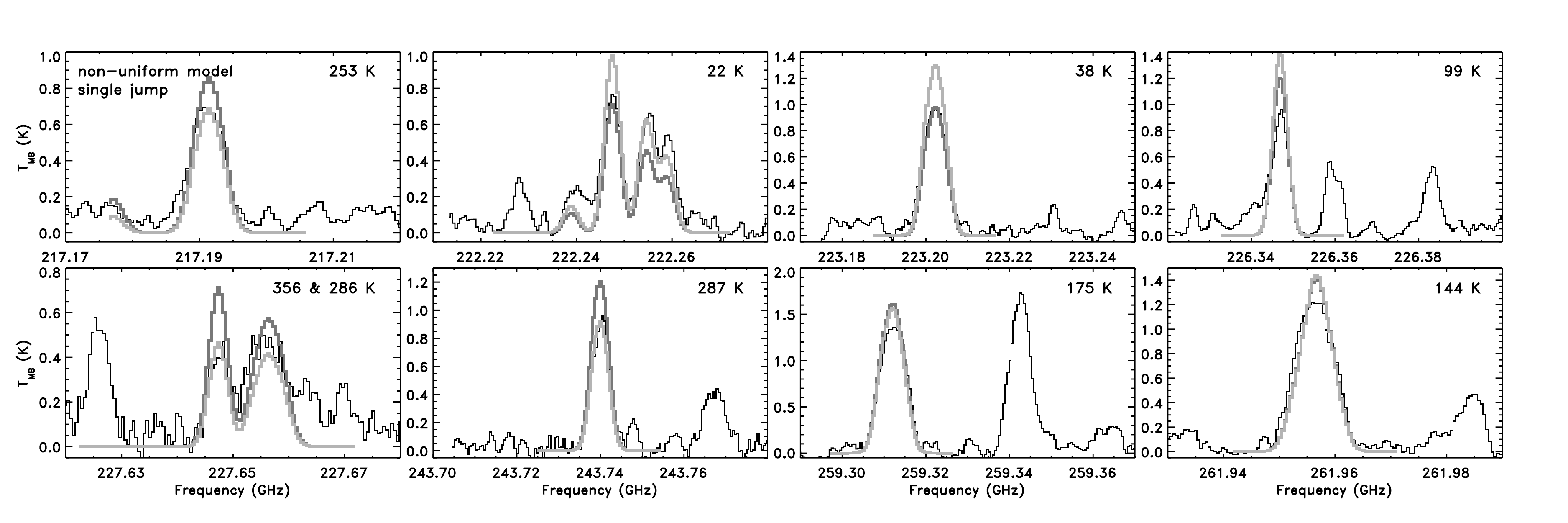}
\caption{Selected emission lines for CH$_3$OCH$_3$, $\varv=0$,
    over-plotted with the best-fit non-uniform {\it LIME} model with a
    single jump of the CH$_3$OCH$_3$ abundance at 70 or 100~K (see
    Table~\ref{chi}). The observed spectrum is displayed in black, the
    model for CH$_3$OCH$_3$, $\varv=0$, with an abundance jump at 70~K
    in light grey and at 100~K in dark grey. The value of $E_{\rm u}$
    for each transition is given in the upper right corner of each
    panel.} \label{lime1jump}
\includegraphics[width=18cm]{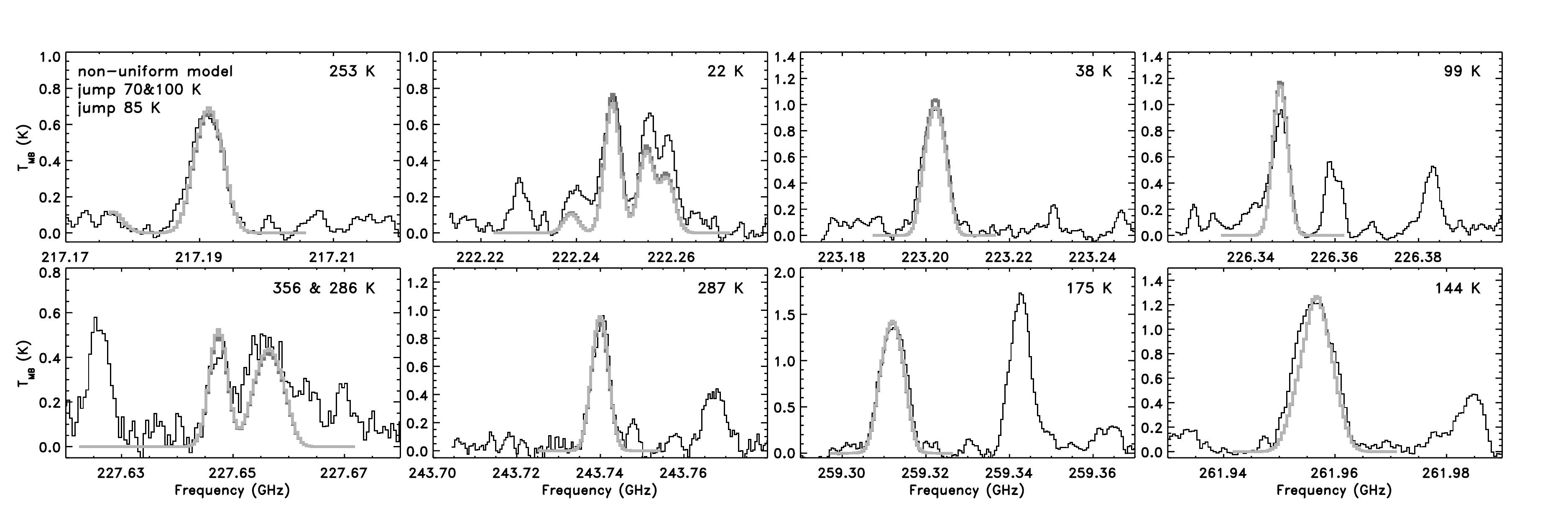}
\caption{Same as Fig.~\ref{lime1jump}, but with two jumps of the 
  CH$_3$OCH$_3$ abundance at 70 and 100~K (light grey) and a single jump of the abundance at 85~K (dark grey). The observed
  spectrum is displayed in black.} \label{limedme}
\includegraphics[width=18cm]{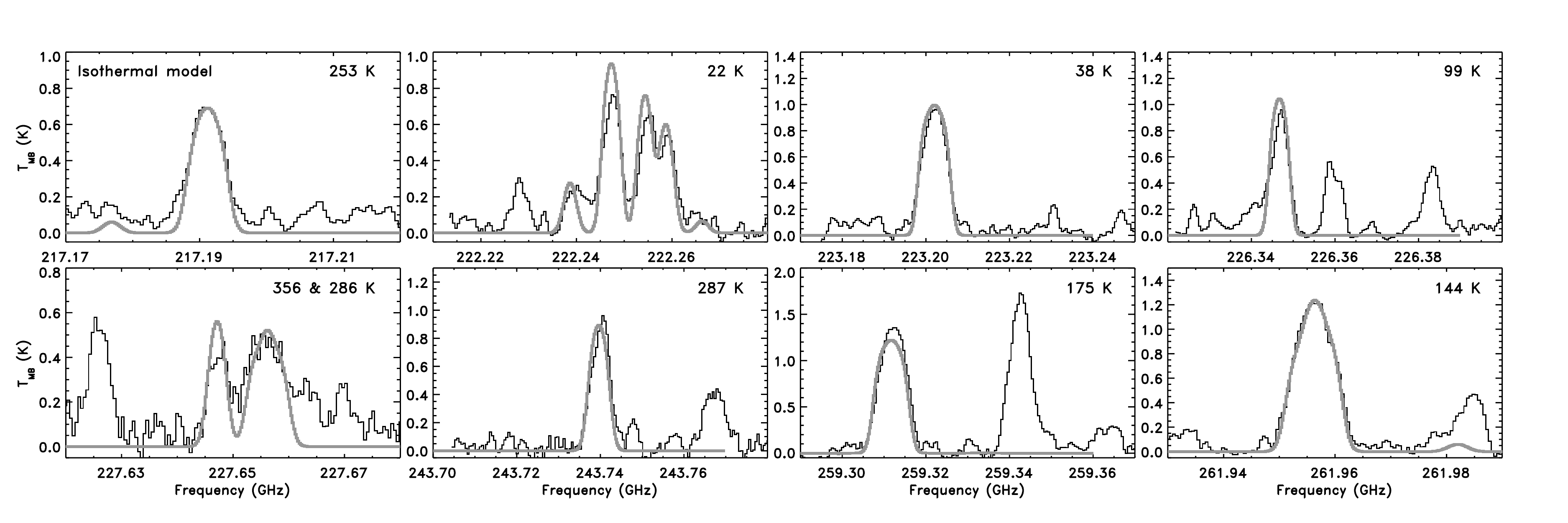}
\caption{Same as Fig.~\ref{limedme}, but with the isothermal {\it myXCLASS}
  model over-plotted in grey (see Table~\ref{sumres} for the
  properties of the two isothermal components).}\label{xclassdme}
\end{figure*}

We have used the density and temperature profile for the source
G327.3-0.6 calculated by \citet{rolffs2011} and modeled the dimethyl
ether emission with the new radiative transfer tool {\it LIME}
\citep{brinch2010}. \citet{rolffs2011} approximate the density and
temperature profile of G327.3-0.6 with a spherical model based on the
870~$\mu$m dust radial profile observed with the Large APEX Bolometer
Camera (LABOCA) in the ATLASGAL project \citep[APEX Telescope Large
Area Survey of the Galaxy,][]{schuller2009}. In the model the source
has a luminosity of 7$\times$10$^4$~L$_\odot$, a temperature of 55~K
at the photospheric radius (i.e., diffusion sets the temperature
gradient in the inner region, and the balance between heating and
cooling sets the temperature gradient in the outer region), and a
density of 3.5$\times$10$^{6}$~cm$^{-3}$. The photospheric radius is
derived by computing the Planck and Rosseland opacities and weighting
them with the Planck function \citep[see][for the
details]{rolffs2011}. The model assumes the gas is static, and has a
turbulent line width, $FWHM$, of 4~km~s$^{-1}$.

The dimethyl ether in the non-uniform density and temperature {\it
  LIME} models is assumed to be in LTE as for the isothermal {\it
  myXCLASS} models presented in Sect.~\ref{groundsec}. The dust
opacity has not been taken into account in the {\it LIME} models, due
to issues of its implementation into the code in combination with
overlapping transitions, which for dimethyl nearly always is the
case. For this reason we have chosen to use only low frequency
dimethyl ether transitions (up to: 261956~MHz) for which the dust is
optically thin (see Sect.~\ref{dust}) and thus does not affect the
model results significantly. Another assumption is that CH$_3$OCH$_3$
evaporates above 70~K and that dimethyl ether's precursor methanol
evaporates at $\sim$100~K (see Sect.~\ref{intro}). The radii at which
these temperatures occur are consistent with those found for the two
components of the isothermal {\it myXCLASS} model described in
Sect.~\ref{groundsec}.

To assess the accuracy of the models, we have performed a reduced
$\chi^2$-analysis for the calculated emission for a set of 7 dimethyl
ether transitions that span a range of temperatures and optical depths
and are known to be free from emission line blends from our survey. In
Table~\ref{chi} we show the best reduced $\chi^2$ for the fits to the
transitions shown in Figs.~\ref{lime1jump}--\ref{xclassdme} for the
non-uniform density and temperature models with the CH$_3$OCH$_3$
abundance jumping at a temperature ranging from 70~K up to
100~K. Results for the best-fit model with a jump at 70~K (light grey)
and 100~K (dark grey) are also shown in Fig.~\ref{lime1jump}, as well
as the model with a jump at both 70 and 100~K (light grey) and a
single jump at 85~K (dark grey) in Fig.~\ref{limedme}. The reduced
$\chi^2$ is calculated for a 10~km~s$^{-1}$ wide window around the
peak position except for the blended features (second and fifth
panel), where the velocity range covered is 8~km~s$^{-1}$ to avoid
including emission from other species. Below the jump temperature the
abundance is set to be negligible (1.0$\times$10$^{-15}$
w.r.t. H$_2$). For each jump temperature, we first tried to identify
around which abundance the reduced $\chi^2$ was lowest and afterwards
ran a set of 4--6 models with abundance steps of 1--2$\times$10$^{-8}$
to identify the best-fit abundance.

As can be seen in Fig.~\ref{lime1jump} the {\it LIME} model, where the
abundance jump for CH$_3$OCH$_3$ occurs at 100~K, fits most emission
lines very well, but seems to over-reproduce the transitions with the
highest excitation energies. In contrast, the model where there is a
single jump at 70~K, has the opposite problem, that the lower
excitation transitions are over-produced, when those at high energies
are well fit. When there are two abundance jumps at 70 and 100~K or
one at an intermediate temperature of 85~K (see Fig.~\ref{limedme})
this systematic mismatch of low or higher excitation transitions is
not observed. The difference between the model with the abundance jump
at 85~K and that with a jump at both 70 and 100~K is, however,
negligible. Both models fit the observed line strength and width very
well. It is only in the second panel for the 22~K lines that some
deviation is seen for two weaker lines, which are slightly
under-produced very likely due to unknown line blends.

Due to the similarities for the non-uniform density {\it LIME} models
with a single or double abundance jump no conclusion about whether
there are one or two jumps in the CH$_3$OCH$_3$ abundance, can be
drawn. Finally, the models suggest that most dimethyl ether emission
arises from a region with temperatures above 100~K, but that some
emission seems to arise from a region with temperatures lower than
100~K.

\begin{table}
  \caption{Results of the fits to a selected number of dimethyl ether transitions for both the non-uniform density and temperature {\it LIME} models as well as the isothermal {\it myXCLASS} models.}\label{chi}
\begin{center}
\begin{tabular}{llll}
  \hline
  \hline
$T_{\rm jump}$& $\chi^2$ & $x$(CH$_3$OCH$_3$)$^a$ & $\frac{\mathcal{N}_{\rm <100~K}{\rm(CH_3OCH_3)}}{\mathcal{N}{\rm (CH_3OCH_3)}}$\\
  K &  & w.r.t. H$_2$  &\\
  \hline
  \multicolumn{4}{l}{Non-uniform density and temperature model}\\
  \hline
\phantom{1}70 & 1.73 & 9$\times$10$^{-8}$ & 0.41\\
\phantom{1}80 & 1.31 & 1.1$\times$10$^{-7}$ & 0.28\\
\phantom{1}85 & 1.26 & 1.4$\times$10$^{-7}$ & 0.21\\
\phantom{1}90 & 1.33 & 1.4$\times$10$^{-7}$ & 0.14\\
100 & 1.37 & 2.0$\times$10$^{-7}$ & 0\phantom{.00}\\
\phantom{1}70 \& 100 & 1.32 & 3$\times$10$^{-8}$ \& 1.4$\times$10$^{-7}$ & 0.05\\
\hline
\multicolumn{4}{l}{Isothermal model}\\
\hline
\phantom{1}90 & 2.08 & 1.7$\times$10$^{-7}$ & 0.14\\
\phantom{1}80 \& 100 & 1.34 & 5.9$\times$10$^{-8}$ \& 3.0$\times$10$^{-7}$ & 0.08\\ 
\hline
\end{tabular}
\end{center}
$^a$The abundance in the outermost part of the envelope, where temperatures are below 70~K, is assumed to be 1.0$\times$10$^{-15}$ with respect to H$_2$, i.e., negligible.
\end{table}

\section{Discussion}
\label{disc}

\subsection{Comparison of an isothermal source model with a
  non-uniform density and temperature model}\label{sm}

In general, we can conclude that isothermal models for the dimethyl
ether emission are in qualitative agreement with the models with a
non-uniform source structure. This is demonstrated by the comparison
of Figs.~\ref{limedme} and \ref{xclassdme}. There are small
differences, i.e., the line wings are better fit with the isothermal
model, whereas the line-centers are often better fit with the
non-uniform source model. To compare the isothermal fit to the
non-uniform spherical model, we have calculated a reduced $\chi^2$ for
the {\it myXCLASS} models with one and two components of the same
lines that are used in the {\it LIME} model. The isothermal model
suggests two abundance jumps for CH$_3$OCH$_3$, as can be seen by the
significantly lower value for $\chi^2$ for the two-component model
(see Sect.~\ref{groundsec}). However, the non-uniform density and
temperature models suggest that two separate abundance jumps are not
needed. This suggests that dimethyl ether traces gas that has a
temperature and density gradient. In a way our two-component
isothermal model can be considered as a ``simple'' non-uniform density
and temperature model. The non-uniform source model is therefore a
useful tool for the interpretation of the emission of CH$_3$OCH$_3$ in
the hot core of G327.3-0.6.

\subsection{Implications for the formation mechanisms of CH$_3$OCH$_3$}\label{constraints-formation} 

The fact that CH$_3$OCH$_3$ is present in regions with a range of
temperatures makes it difficult to separate the contribution of grain
surface formation or gas phase formation and determine which is
dominant. If we assume that dimethyl ether with excitation
temperatures below 100~K is due to solid state formation, then it is
possible to estimate the lower limit to the fraction of CH$_3$OCH$_3$
that is formed on icy grains. The reason we can only deduce lower
limits is, that the dimethyl ether present at higher temperatures may
also originate from solid state chemistry. A laboratory study by
\citet{collings2004} namely shows that many molecular species can be
trapped in ices up to the H$_2$O evaporation temperature. For species
with similar binding energies as CH$_3$OCH$_3$, about 50\% remains
trapped in the ice until the H$_2$O evaporates when it is co-deposited
with water. This fraction is lower for layered ices, where the studied
species was deposited on top of a water ice layer. Unfortunately,
CH$_3$OCH$_3$ was not one of the species studied by
\citet{collings2004}, but it is reasonable to assume that something
similar would occur for dimethyl ether and that some fraction of the
dimethyl ether gas at temperatures above 100~K is due to CH$_3$OCH$_3$
co-desorbing with H$_2$O.

In Table~\ref{chi} we show the fractional contribution of the dimethyl
ether containing gas that has a temperature between 70 and 100~K,
$\frac{\mathcal{N}_{\rm <100~K}{\rm(CH_3OCH_3)}}{\mathcal{N}{\rm
    (CH_3OCH_3)}}$. $\mathcal{N}{\rm (CH_3OCH_3)}$ is defined as the
total number of dimethyl ether molecules in the model, and is derived
by integrating over the radial density profile combined with the
radial abundance dependence for dimethyl ether. $\mathcal{N}_{\rm
  <100~K}{\rm(CH_3OCH_3)}$ is calculated in a similar way, but only by
integrating over the part of the envelope that is below 100~K. If we
consider the model with the abundance jump at 85~K, the fraction of
dimethyl ether present in gas between 70 and 100~K is 21\%. This would
suggest that at least 21\% is formed on icy grains. If we assume that
the two-jump model is the most extreme case, since, as mentioned
earlier in this section, the 100~K systematically under-reproduces all
the lower excitation lines, then the models suggest that the lower
limit to the grain surface contribution is 5\% and conversely, this
means that the upper limit to the gas phase contribution is 95\%.

An alternative formation mechanism for low-temperature gas-phase
dimethyl ether is that a small fraction of solid CH$_3$OH evaporates
from the ices at low temperatures and gives rise to a small amount of
CH$_3$OCH$_3$ that is formed in the cooler gas phase. For
$^{13}$CH$_3$OH we have performed a similar analysis as described for
dimethyl ether in this paper, i.e. we modeled its emission with both
isothermal and a non-uniform density and temperature models. The
best-fit abundances found for $^{13}$CH$_3$OH in non-uniform density
and temperature models are 2($\pm$0.5)$\times$10$^{-11}$ and
4.5($\pm$0.5)$\times$10$^{-8}$, below and above 100~K,
respectively. When we assume the $^{12}$C/$^{13}$C ratio to be 53 the
average found by \citet{wilson1994} for the 4~kpc molecular ring, the
CH$_3$OH abundance below 100~K in G327.3-0.6 would be
$\sim$1$\times$10$^{-9}$ and $\sim$2$\times$10$^{-6}$ above
100~K. This is consistent with models of both low and high-mass
star-forming regions where the abundance of CH$_3$OH is on the order
of 10$^{-10}$--10$^{-9}$ below 90--100~K
\citep{schoier2006,maret2005,vdtak2000b}. Since radiative dissociation
experiments performed with CD$_3$OCD$_4^+$ by \citet{hamberg2010} only
results in the formation of deuterated dimethyl ether in 7\% of the
reactions, and not all CH$_3$OH will react to form CH$_3$OCH$_4^+$, we
expect the CH$_3$OCH$_3$/CH$_3$OH ratio to be 0.07 at most, but
probably much lower. The gas phase abundances of dimethyl ether formed
at low temperatures is therefore likely not larger than
7$\times$10$^{-11}$. Additionally, CH$_3$OCH$_3$ itself may also
desorb directly through non-thermal mechanisms. In theoretical models
of grain-surface formation the solid state CH$_3$OCH$_3$/CH$_3$OH
ratio is on the order of 10$^{-3}$--10$^{-4}$ \citep{garrod2008}. The
gas phase abundance of CH$_3$OCH$_3$ that could be due to non-thermal
desorption is thus expected to be $\sim$10$^{-13}$, which is lower
than the CH$_3$OCH$_3$ abundance that could be formed from
non-thermally desorbed CH$_3$OH. Since abundances of $\sim$10$^{-8}$
for dimethyl ether are needed in the outer regions both in the
isothermal case and for the non-uniform source models for the
G327.3-0.6 high-mass star-forming region, the major contributor to
cooler dimethyl ether gas cannot be direct non-thermally desorbed
CH$_3$OCH$_3$ or gas phase formation through the non-thermal
evaporation of CH$_3$OH. CH$_3$OCH$_3$ that desorbs thermally from icy
grain surfaces is therefore most likely responsible for the
CH$_3$OCH$_3$ emission below 100~K. Interestingly, the
CH$_3$OCH$_3$/CH$_3$OH ratio above 100~K is in the 5--10\% range for
our non-spherical source models, which is close to the upper limit for
the ratio expected from gas phase formation mechanism, underlining
that although the data suggests that there is some grain-surface
formation of dimethyl ether, the gas phase mechanism likely is
dominant.

In summary, our non-uniform density and temperature {\it LIME} models
of the emission of dimethyl ether, give qualitatively the same result
as the isothermal {\it myXCLASS} model. They namely suggest that most
emission of dimethyl ether arises from regions with temperatures of
100~K and higher, but that some fraction of dimethyl ether emission is
present at lower temperatures. The most likely origin of this low
temperature component is formation of dimethyl ether in the solid
state.

\section{Summary and conclusions}\label{conclusion}

In this paper we have analyzed the rotational emission of
CH$_3$OCH$_3$ in its ground state, $\varv=0$, and the torsionally
excited states $\varv_{11}=1$ and $\varv_{15}=1$ for the source
G327.3-0.6 observed with the Atacama Pathfinder EXperiment (APEX). The
data have been modeled with an isothermal model as well as a
radiative transfer model with a non-uniform spherical density and
temperature structure. The main conclusions are:

\begin{itemize}
\item Many transitions from the ground state, $\varv=0$, of dimethyl
  ether are detected toward the high-mass star forming region
  G327.3-0.6. The emission can be very well described by a
  two-component isothermal model with two abundance jumps at 80~K and
  100~K. Since dimethyl ether has both optically thin and moderately
  optically thick lines, it is a very suitable diagnostic for dense
  and fairly warm regions.
\item A variety of lines from the $\varv_{11}=1$ and $\varv_{15}=1$
  torsionally excited states of dimethyl ether are detected here for
  the first time. The emission can be well described by the same model
  as for the ground state. Due to the large line-strengths these
  states should also easily be detectable in other line-rich hot core
  sources, in particular with new upcoming interferometers such as the
  Atacama Large Millimeter Array (ALMA).
\item Radiative transfer models with a non-uniform density and
  temperature structure show that the emission can be very well fit by
  an abundance jump in the dimethyl ether emission at either 85~K or
  by two abundance jumps at 70 and 100~K. In contrast two components
  are needed to accurately reproduce the emission with an isothermal
  model. We therefore conclude that the non-uniform density and
  temperature models suggest that dimethyl ether emission arises from
  an extended region with a significant density and temperature
  structure, but not necessarily has two abundance components. We
  estimate that the lower limit for the solid state vs. gas phase
  contribution is 5\%. However, the large abundance ratio
  CH$_3$OCH$_3$/CH$_3$OH suggest that the gas phase formation
  mechanism is likely the major contributor.
\end{itemize}

The presented study shows the potential for these kind of surveys to
constrain the chemical structure of high-mass protostellar hot
cores. The study of CH$_3$O$^{13}$CH$_3$ could potentially give
additional clues to the formation of dimethyl ether. The
$^{12}$C/$^{13}$C ratio namely has the potential to distinguish
between grain surface formation and gas phase formation of dimethyl
ether \citep{charnley2004a}. This would make it possible to not just
improve our understanding of the formation routes of dimethyl ether,
but also address whether isotopologues in general can be used as
important tracers of the chemistry in star-forming regions.

In addition, it would be interesting to perform similar analyses for a
variety of complex organic species. This will make it possible to
compare different molecular species with each other as well as to
astrochemical models. Specifically, a strong correlation between the
abundances of different complex organic species in large sample of
star-forming regions, such as performed by \citet{bisschop2007a} may
tell us whether two complex organic molecules likely are chemically
related. In the longer term, high resolution interferometric
measurements, in particular with ALMA, are key to understanding the
origin of CH$_3$OCH$_3$ and other complex organics in star-forming
regions. The higher spatial resolution will place better constraints
on the physical structure of the source and will for example show
whether the emission is clumpy or non-spherical. Also, such
observations will show how the absolute abundances vary with
temperature and thus reveal whether a given species comes off dust
grains directly at its evaporation temperature - or perhaps is formed
only at even higher temperatures in the gas-phase. From this it is
possible to conclude what the relative importance of solid state and
gas phase formation mechanisms are and for which key reactions
laboratory investigations and quantum chemical calculations are most
needed.

\begin{acknowledgements}
  The research of SEB was supported by a Rubicon grant from the
  Netherlands Organization for Scientific Research and a grant from
  Instrument center for Danish Astrophysics. The research in
  Copenhagen was furthermore supported by Centre for Star and Planet
  Formation, which is funded by the Danish National Research
  Foundation and the University of Copenhagen’s program of excellence
  and a Junior Group Leader Fellowship to JKJ from the Lundbeck
  foundation. H.S.P.M. is very grateful to the Bundesministerium f\"ur
  Bildung und Forschung (BMBF) for financial support aimed at
  maintaining the Cologne Database for Molecular Spectroscopy,
  CDMS. This support has been administered by the Deutsches Zentrum
  f\"ur Luft- und Raumfahrt (DLR). Laboratory work on complex
  molecules in funded by DFG within CRC956. We would also like to
  thank an anonymous referee and the editor Malcolm Walmsley for
  constructive comments on this paper.
\end{acknowledgements}

\begin{appendix}
  \section{Assignments of CH$_3$OCH$_3$,  $\varv=0$,  $\varv_{11}=1$, and
    $\varv_{15}=1$ for G327.3-0.6 in all observed frequency
    windows.}\label{ap1}

  In this appendix we show the line parameters for all the unblended
  lines that have been detected for CH$_3$OCH$_3$, $\varv=0$
  (Table~\ref{detfreqv0}), $\varv_{11}=1$ (Table~\ref{detfreqv11}),
  and $\varv_{15}=1$ (Table~\ref{detfreqv15}). For questions about the
  full line-list please contact the authors. For all transitions
  frequencies, quantum numbers and the energies of the upper state
  ($E_{\rm u}$) are given as well as the measured integrated line
  intensities ($\int T_{\rm MB}dv$) and model line-intensities from
  the isothermal {\it myXCLASS} model for the dimethyl ether emission
  ($I_{\rm model}$). Most features are actually composed of a set of
  multiple transitions, and we therefore refer to \citet{endres2009}
  and Endres et al. (in prep.) for the values of $\mu^2S$ for the
  individual transitions.

\onecolumn

\begin{longtable}{lllll}
\caption[Detections of CH$_3$OCH$_3$, $\varv=0$ transitions found to be free of emission from other species towards G327.3-0.6.]{Detections of CH$_3$OCH$_3$, $\varv=0$ towards G327.3-0.6.}\label{detfreqv0}\\
\hline
\hline
Frequency & Transition & $E_{\rm u}$ & $\int T_{\rm MB}dv$ & $I_{\rm model}$ \\
(MHz) & & (K) & (K~km~s$^{-1}$) & (K~km~s$^{-1}$)\\
\hline
\endfirsthead
\multicolumn{5}{l}
{{\tablename\ \thetable{} --Continued from previous page}} \\
\hline
\hline
Frequency & Transition & $E_{\rm u}$ & $\int T_{\rm MB}dv$ & $I_{\rm model}$ \\
(MHz) & & (K) & (K~km~s$^{-1}$) & (K~km~s$^{-1}$)\\
\hline
\endhead
\hline \multicolumn{5}{|c|}{{Continued on next page}} \\ \hline
\endfoot
\hline \hline
\endlastfoot
213453.7 &       21$_{       3,       1,      20,       3/      5}$ --       21$_{       1,       0,      21,       3/      5}$ & 212.92 & \phantom{1}1.60 & \phantom{1}1.30\\
213457.7 &       21$_{       3,       1,      20,       1}$ --       21$_{       1,       0,      21,       1}$ & 212.92 & \phantom{1}2.40 & \phantom{1}2.19\\
213461.7 &       21$_{       3,       1,      20,       0}$ --       21$_{       1,       0,      21,       0}$ & 212.92 & \phantom{1}1.42 & \phantom{1}1.24\\
214470.7 &       28$_{      11,       5,      23,       0/      1/      3/      5}$ --       28$_{       9,       4,      24,       0/      1/      3/      5}$ & 406.35 & \phantom{1}2.54 & \phantom{1}1.76\\
217176.9 &       36$_{       9,       4,      32,       0/      1/      3/      5}$ --       36$_{       7,       3,      33,       0/      1/      3/      5}$ & 638.63 & \phantom{1}0.18 & \phantom{1}0.20\\
217191.0 &       22$_{       8,       4,      19,       0/      1/      3/      5}$ --       22$_{       6,       3,      20,       0/      1/      3/      5}$ & 253.41 & \phantom{1}4.75 & \phantom{1}5.44\\
218491.3 &       23$_{       6,       3,      21,       0/      1/      3/      5}$ --       23$_{       4,       2,      22,       0/      1/      3/      5}$ & 263.83 & \phantom{1}3.45 & \phantom{1}3.86\\
220893.0 &       23$_{       8,       4,      20,       0/      1/      3/      5}$ --       23$_{       6,       3,      21,       0/      1/      3/      5}$ & 274.44 & \phantom{1}4.44 & \phantom{1}4.16\\
\hline
\end{longtable}

\begin{longtable}{lllll}
  \caption[Detections of CH$_3$OCH$_3$, $\varv_{11}=1$, transitions found to be free of emission from other species towards G327.3-0.6.]{Detections of CH$_3$OCH$_3$, $\varv_{11}=1$, transitions found to be free of emission from other species towards G327.3-0.6.}\label{detfreqv11}\\
\hline
\hline
Frequency & Transition & $E_{\rm u}$ & $\int T_{\rm MB}dv$ & $I_{\rm model}$ \\
(MHz) & & (K) & (K~km~s$^{-1}$) & (K~km~s$^{-1}$)\\
\hline
\endfirsthead
\multicolumn{5}{l}
{{\tablename\ \thetable{} --Continued from previous page}} \\
\hline
\hline
Frequency & Transition & $E_{\rm u}$ & $\int T_{\rm MB}dv$ & $I_{\rm model}$ \\
(MHz) & & (K) & (K~km~s$^{-1}$) & (K~km~s$^{-1}$)\\
\hline
\endhead
\hline \multicolumn{5}{|c|}{{Continued on next page}} \\ \hline
\endfoot
\hline \hline
\endlastfoot
225226.4 & 12$_{1,12,0}$ -- 11$_{0,11,0}$ & 355.67 & \phantom{1}1.15 & \phantom{1}0.68\\
225245.0 & 12$_{1,12,3/5}$ -- 11$_{0,11,3/5}$ & 355.65 & \phantom{1}1.02 & \phantom{1}0.69\\
241538.1 & 13$_{1,13,3/5}$ -- 12$_{0,12,3/5}$ & 366.97 & \phantom{1}0.83 & \phantom{1}0.50\\
250263.7 & 14$_{0,14,5/3/1/0}$ -- 13$_{1,13,5/3/1/0}$ & 378.98 & \phantom{1}3.85 & \phantom{1}2.20\\
252067.7 & 21$_{5,16,1}$ -- 21$_{4,17,1}$ & 531.33 & \phantom{1}1.36 & \phantom{1}0.34\\
254754.5 & 20$_{5,15,1}$ -- 20$_{4,16,1}$ & 512.19 & \phantom{1}0.85 & \phantom{1}0.40\\
255091.1 & 18$_{2,16,1}$ -- 17$_{3,15,1}$ & 450.02 & \phantom{1}0.54 & \phantom{1}0.28\\
256924.9 & 19$_{5,14,1}$ -- 19$_{4,15,1}$ & 493.96 & \phantom{1}0.90 & \phantom{1}0.44\\
\hline
\end{longtable}

\begin{longtable}{lllll}
  \caption[Detections of CH$_3$OCH$_3$, $\varv_{15}=1$, found to be free of emission from other species towards G327.3-0.6.]{Detections of CH$_3$OCH$_3$, $\varv_{15}=1$, found to be free of emission from other species towards G327.3-0.6.}\label{detfreqv15}\\
\hline
\hline
Frequency & Transition & $E_{\rm u}$ & $\int T_{\rm MB}dv$ & $I_{\rm model}$ \\
(MHz) & & (K) & (K~km~s$^{-1}$) & (K~km~s$^{-1}$)\\
\hline
\endfirsthead
\multicolumn{5}{l}
{{\tablename\ \thetable{} --Continued from previous page}} \\
\hline
\hline
Frequency & Transition & $E_{\rm u}$ & $\int T_{\rm MB}dv$ & $I_{\rm model}$ \\
(MHz) & & (K) & (K~km~s$^{-1}$) & (K~km~s$^{-1}$)\\
\hline
\endhead
\hline \multicolumn{5}{|c|}{{Continued on next page}} \\ \hline
\endfoot
\hline \hline
\endlastfoot
225482.3 & 12$_{1,12,0}$ -- 11$_{0,11,0}$ & 416.36 & \phantom{1}0.54 & \phantom{1}0.18\\
225491.5 & 12$_{1,12,1}$ -- 11$_{0,11,1}$ & 416.35 & \phantom{1}0.96 & \phantom{1}0.48\\
250026.9 & 14$_{0,14,5/3/1/0}$ -- 13$_{1,13,5/3/1/0}$ & 439.66 & \phantom{1}2.58 & \phantom{1}1.37\\
260934.8 & 19$_{5,15,1}$ -- 19$_{4,16,1}$ & 554.36 & \phantom{1}0.74 & \phantom{1}0.18\\
263123.2 & 13$_{5,8,1}$ -- 13$_{4,9,1}$ & 464.45 & \phantom{1}0.27 & \phantom{1}0.28\\
263323.6 & 12$_{5,8,1}$ -- 12$_{4,9,1}$ & 452.67 & \phantom{1}0.47 & \phantom{1}0.31\\
263389.0 & 12$_{5,7,1/3}$ -- 12$_{4,8,1/3}$ & 452.65 & \phantom{1}1.25 & \phantom{1}0.41\\
263736.8 & 10$_{5,6,1}$ -- 10$_{4,7,1}$ & 431.81 & \phantom{1}0.85 & \phantom{1}0.32\\
\hline
\end{longtable}

\end{appendix}
\twocolumn
\end{document}